\documentstyle[12pt]{article}
\def\ie{{\em i.e.}}
\def\eg{{\em e.g.}}

\def\beq{\begin{equation}}
\def\eeq{\end{equation}}

\catcode`\@=11 

\def\lsim{\mathrel{\mathpalette\@versim<}}
\def\gsim{\mathrel{\mathpalette\@versim>}}
\def\@versim#1#2{\vcenter{\offinterlineskip
    \ialign{$\m@th#1\hfil##\hfil$\crcr#2\crcr\sim\crcr } }}

\def\JL{J. L. Lopez}
\def\DVN{D. V. Nanopoulos}
\def\AZ{A. Zichichi}

\def\t1{{\tilde 1}}

\def\GeV{\,{\rm GeV}}

\def\to{\rightarrow}

\def\NPB#1#2#3{Nucl. Phys. B {\bf#1} (19#2) #3}
\def\PLB#1#2#3{Phys. Lett. B {\bf#1} (19#2) #3}

\def\PRD#1#2#3{Phys. Rev. D {\bf#1} (19#2) #3}

\def\IJMP#1#2#3{Int. J. Mod. Phys. A {\bf#1} (19#2) #3}

\begin{document}
\begin{flushright}
\baselineskip=12pt
{CERN-TH.7553/95}\\
{CTP-TAMU-67/94}\\
{ACT-24/94}\\
{hep-ph/9501258}\\
\end{flushright}

\begin{center}
\vglue 1cm
{\Huge\bf R$_{b}$ in supergravity models\\}
\vglue 2cm
{XU~WANG$^{1,2}$, JORGE L. LOPEZ$^{1,2}$, and D. V. NANOPOULOS$^{1,2,3}$\\}
\vglue 1cm
{\em $^{1}$Center for Theoretical Physics, Department of Physics, Texas A\&M
University\\}
{\em College Station, TX 77843--4242, USA\\}
{\em $^{2}$Astroparticle Physics Group, Houston Advanced Research Center
(HARC)\\}
{\em The Mitchell Campus, The Woodlands, TX 77381, USA\\}
{\em $^{3}$CERN Theory Division, 1211 Geneva 23, Switzerland\\}
\baselineskip=12pt
\end{center}

\vglue 1cm
\begin{abstract}
We compute the supersymmetric contribution to $R_{b}\equiv \Gamma (Z\to
b{\bar b})/\Gamma (Z\to {\rm hadrons})$ in a variety of supergravity models.
We find $R^{\rm susy}_b\lsim0.0004$, which does not shift significantly
the Standard Model prediction ($R^{\rm SM}_b=0.2162$ for $m_t=160\GeV$).
An improvement in experimental precision by a factor of four would be
required to be sensitive to such an effect.
\end{abstract}

\vspace{3cm}
\begin{flushleft}
\baselineskip=12pt
{CERN-TH.7553/95}\\
{CTP-TAMU-67/94}\\
{ACT-24/94}\\
January 1995
\end{flushleft}
\newpage

\setcounter{page}{1}
\pagestyle{plain}
\baselineskip=14pt

Precision tests of the electroweak interactions at LEP have provided the
most sensitive checks of the Standard Model of particle physics. The pattern
that has emerged is that of consistent agreement with the Standard Model
predictions. This pattern seems to have so far only one apparently dissonant
note, namely in the measurement of the ratio $R_b\equiv \Gamma(Z\to b{\bar
b})/\Gamma(Z\to{\rm hadrons})$, where the latest global fit to the LEP data
($0.2192\pm0.0018$ \cite{Rbexp}) lies about two standard deviations above the
one-loop Standard Model prediction \cite{RbSM} for all preferred values of the
top-quark mass (\eg, $R_b^{\rm SM}=0.2162$ for $m_t=160\GeV$). Further
experimental statistics will reveal whether this is indeed a breakdown of the
Standard Model. In the meantime, it is important to explore what new
contributions to $R_b$ are expected in models of new physics, such as
supersymmetry.

The study of supersymmetric contributions to $\Gamma(Z\to b\bar b)$ has
proceeded in two phases. Originally the quantity $\epsilon_b$ \cite{ABC-Rb} was
defined as an extension of the $\epsilon_{1,2,3}$ scheme \cite{ABC} for
model-independent fits to the electroweak data. More recently it has become
apparent that the ratio $R_b$ is more directly calculable \cite{BF,KKW} and
readily measurable. It has been made apparent \cite{KKW} that supersymmetric
contributions to $R_b$ are not likely to increase the total predicted value for
$R_b$ in any significant manner, as long as typical assumptions about unified
supergravity models are made. On the other hand, if these assumptions are
relaxed, it is possible for supersymmetry to make a significant contribution to
$R_b$, if certain conditions on the low-energy supersymmetric spectrum are
satisfied: the lightest chargino and lightest top-squark should be below
$100\GeV$, with the chargino being mostly Higgsino and the top-squark mostly
right-handed. One should keep in mind that novel supersymmetry breaking
scenarios, such as those arising from string models, may provide such otherwise
ad-hoc conditions. In any event, for the purposes of this paper we assume that
the experimental data will settle at values for $R_b$ which do not require
extreme choices of the supersymmetry parameters for its explanation, and yet
still allow for a discrimination among the various supergravity models on the
basis of their prediction for $R_b$. If this assumption turns out to be
invalid, all models discussed in this paper (as well as the Standard Model)
would be seriously disfavored.

We consider unified supergravity models with universal soft supersymmetry
breaking at the unification scale, and radiative electroweak symmetry breaking
(enforced using the one-loop effective potential) at the weak scale
\cite{reviews}. These constraints reduce the number of parameters needed to
describe the models to four, which can be taken to be $m_{\chi^\pm_1},
\xi_0\equiv m_0/m_{1/2},\xi_A\equiv A/m_{1/2},\tan\beta$, with a specified
value for the top-quark mass ($m_t$). In what follows we take $m_t^{\rm
pole}=160\GeV$ which is the central value obtained in fits to all electroweak
and Tevatron data in the context of supersymmetric models~\cite{EFL}. Among
these four-parameter supersymmetric models we consider generic models
with continuous values of $m_{\chi^\pm_1}$ and discrete choices for the other
three parameters:
\begin{equation}
\tan\beta=2,10,20\ ;\qquad \xi_0=0,1,2,5\ ;\qquad \xi_A=0\ .
\label{generic}
\end{equation}
The choices of $\tan\beta$ are representative; higher values of $\tan\beta$ are
likely to yield values of $B(b\to s\gamma)$ in conflict with present
experimental limits \cite{LargeTanB}. The choices of $\xi_0$ correspond to
$m_{\tilde q}\approx(0.8,0.9,1.1,1.9)m_{\tilde g}$. The choice of $A$ has
little impact on the results.
We also consider the case of no-scale $SU(5)\times U(1)$ supergravity
\cite{reviews}. In this class of models the supersymmetry breaking parameters
are related in a string-inspired way. In the two-parameter {\em moduli}
scenario $\xi_0=\xi_A=0$ \cite{LNZI}, whereas in the {\em dilaton} scenario
$\xi_0={1\over\sqrt{3}},\ \xi_A=-1$ \cite{LNZII}.
A series of experimental constraints and predictions for these models have been
given in Ref.~\cite{Easpects}. In particular, the issue of precision
electroweak tests in this class of models has been addressed in
Refs.~\cite{ewcorr,eps1-epsb}.

Besides the one-loop Standard Model contributions to $R_b$, in supersymmetric
models there are four new diagrams, as follows:
\begin{itemize}
\item The charged-Higgs--top-quark loop, depends on the charged Higgs mass
and the $t-b-H^\pm$ coupling. For a left-handed $b$ quark the coupling is
$\propto m_{t}/\tan\beta$ whereas for a right-handed $b$ quark it is $\propto
m_{b}\tan\beta$. Therefore, for small\footnote{The radiative electroweak
symmetry breaking mechanism requires $\tan\beta>1$.}  (large) $\tan\beta$ left-
(right)-handed $b$-quark production is dominant. (For $\tan\beta\gg1$, the
value of $m_b$ impacts the contribution significantly.)
It has been shown that the $H^\pm-t$ contribution is always negative
\cite{hollik}, a fact which makes the prediction for $R_b$ in two Higgs-doublet
models always in worse agreement with experiment.
\item The chargino--stop loop, is the supersymmetric counterpart of the
$H^\pm$--$t$ loop discussed above. The chargino mass eigenstate is a mixture of
(charged) Higgsino and wino, and the coupling strength is a complicated matter
now because it involves the stop mixing matrix and the chargino mixing matrix.
However, because only the Higgsino admixture in the chargino eigenstate has a
Yukawa coupling to the $t$--$b$ doublet, generally speaking a light
chargino with a significant Higgsino component, and a light stop
with a significant right-handed component are required for this diagram to make
a non-negligible contribution to $R_b$, as pointed out in Ref.~\cite{KKW}.
\item The neutralino--sbottom loop, is the supersymmetric counterpart of the
neutral-Higgs--bottom-quark loop. The coupling strength of
$\chi_1^{0}-{\tilde b}-b$ is also rather complicated, since it involves the
sbottom mixing, the neutralino mixing, and their masses. However, it can be
non-negligible since it is proportional to $m_b\tan\beta$ for the left-handed
$b$-quark. Therefore in the high $\tan\beta$ region we have to include this
contribution.\footnote{This term was not included in our previous study in
terms of the parameter $\epsilon_b$ \cite{eps1-epsb}, although it was pointed
out that its effects were non-negligible for $\tan\beta\gg1$.}
\item The neutral-Higgs--bottom-quark loop, involves the three neutral scalars,
two CP-even ($h, H$) and one CP-odd ($A$). For the $h\,(H)$ neutral Higgs
boson the coupling to the bottom quark is $\propto m_b \sin\alpha (\cos\alpha)$
which in the absence of a $\tan\beta$ enhancement makes its contribution
negligible. For the $A$ Higgs boson, the coupling to $b\bar b$ is
$\propto\tan\beta$ and the $A$-dependent contribution can be large and positive
if $m_A\lsim90\GeV$ and $\tan\beta\gsim30$ \cite{hollik}. Since we restrict
ourselves to $\tan\beta\lsim20$, and $m_A\gsim100\GeV$ in these models, this
contribution is neglected in what follows.
\end{itemize}

\begin{figure}[p]
\vspace{6in}
\includegraphics{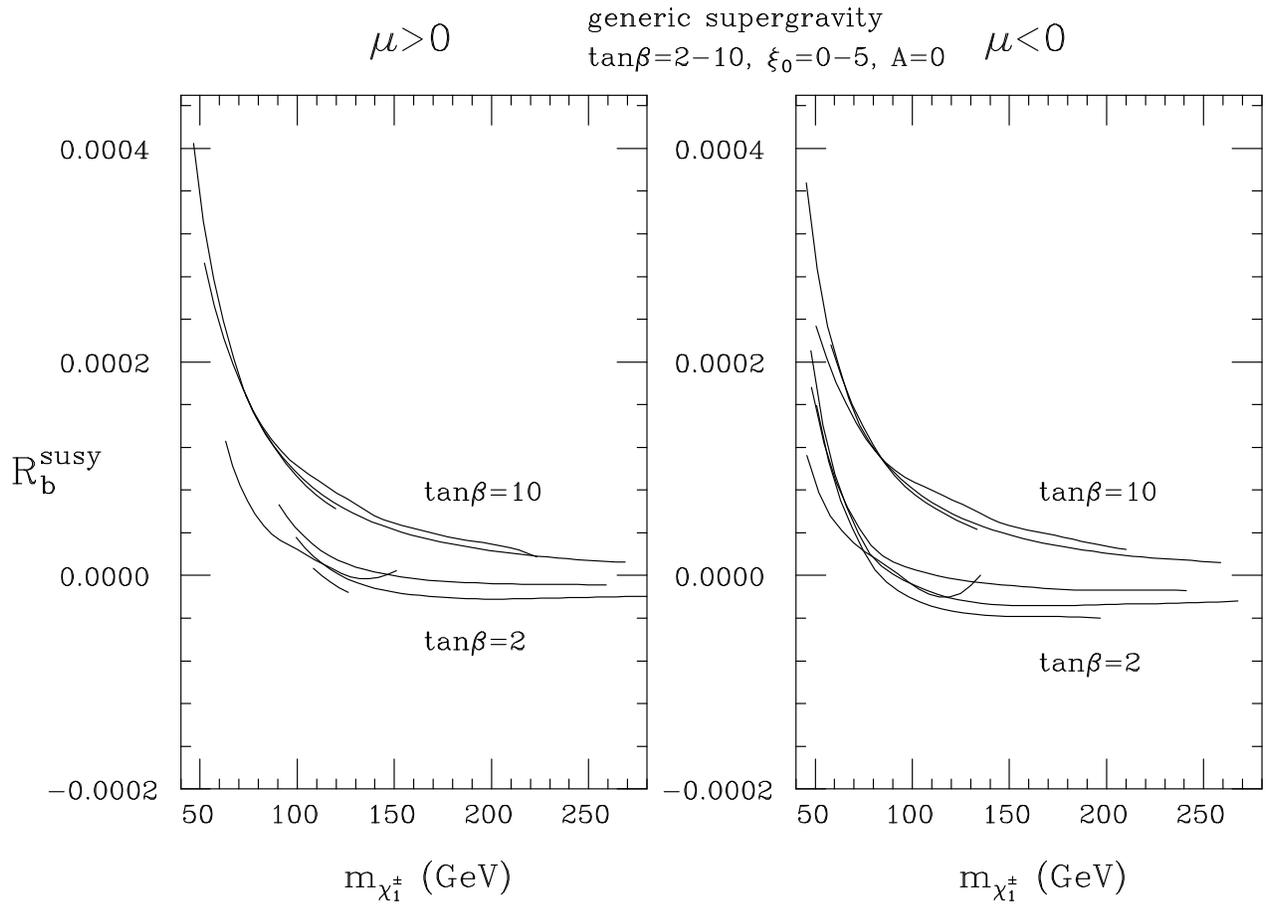}
\caption{The supersymmetric contribution to $R_b$ as a function of the chargino
mass in generic supergravity models with $\tan\beta=2,10$, $\xi_0=0-5$, and
$A=0$. Curves for other values of $\tan\beta$ fall between the two sets shown.}
\label{Rb_SSM}
\end{figure}

Our computations of $R_b^{\rm susy}$ have been performed using the expressions
given in Ref.~\cite{KKW}. Even though these formulas are given explicitly, the
details are quite complicated by the presence of various Passarino-Veltman
loop functions. As a check of our calculations, we have verified numerically
that the results are independent of the unphysical renormalization scale that
appears in the formulas. The predictions for $R^{\rm susy}_b$ in the generic
supergravity models are shown in Fig.~\ref{Rb_SSM}. Only curves for
$\tan\beta=2,10$ are shown, since the corresponding sets of curves for other
values of $\tan\beta$ fall between these two sets of curves. Moreover, in the
interest of brevity we do not show explictly the results in $SU(5)\times
U(1)$ supergravity since for the same values of the parameters the predictions
differ little from those in the generic models.

In almost all cases the largest positive contribution to $R_b^{\rm susy}$ comes
from the chargino--stop loop. As expected, the largest contribution
from this diagram happens for points with the lightest chargino masses (which
correspond to the lightest ${\tilde t}_1$ masses) since supersymmetry is a
decoupling theory. However, even the largest value $(\approx 10^{-4})$ is still
very small compared with the largest possible result in a generic low-energy
supersymmetric model \cite{KKW}. The reason for this is that while
the smallest possible chargino and a stop masses are required for an enhanced
contribution,  it is also necessary that the chargino has a significant
Higgsino component and that the stop
be mostly right-handed. The latter requirement is in fact attainable in these
models (\ie, the stop mixing angle is not small), but the former is not.
Indeed, $R_b^{\rm susy}\propto|V_{12}|^2$, where $V_{12}$ is the (12) element
of the chargino mixing matrix $V$, which does not exceed $\approx0.3$, since
for light charginos $|\mu|\gg M_2$ which makes the lightest chargino mainly a
wino instead of a Higgsino. The charged-Higgs--top-quark loop is
always negative, and is enhanced for either small or large values of
$\tan\beta$. The neutralino--sbottom contribution is almost always smaller than
the chargino--stop contribution and not of definite sign.

We conclude that the presently known supersymmetry breaking scenarios in the
context of supergravity shift only slightly the Standard Model prediction for
$R_b$, and would require an improvement in experimental sensitivity by a factor
of four to be observable. Also, if the experimental value of $R_b$ remains
essentially unchanged over time, the models explored in this paper and the
Standard Model as well, would fall into disfavor. If the experimental value
changes in the direction of the Standard Model prediction, our calculations
should help in discriminating among the various supergravity models on the
basis of their prediction for $R_b$. It is also possible that new scenarios
for supersymmetry breaking may entail low-energy spectra which satisfy the
necessary conditions for an enhanced supersymmetric contribution to $R_b$.

\section*{Acknowledgements}
This work has been supported in part by DOE grant DE-FG05-91-ER-40633. The
work of X. W. has been supported by the World Laboratory. We would like to
thank Chris Kolda for helpful communications. X. W. also would like to thank
J.~T.~Liu for helpful discussions.

\clearpage


\baselineskip=13pt
\begin{thebibliography}{99}
\bibitem{Rbexp}The LEP Collaborations, CERN/PPE/94-187 (November 1994).
\bibitem{RbSM} J. Bernabeu, A. Pich, and A. Santamaria, \PLB{200}{88}{569};
 W. Beenaker and W. Hollik, Z. Phys. C40, 141(1988); A. Akhundov, D. Bardin,
and T. Riemann, \NPB{276}{86}{1}; F. Boudjema, A. Djouadi, and C. Verzegnassi,
\PLB{238}{90}{423}; A. Blondel and C. Verzegnassi, \PLB{311}{93}{346}.
\bibitem{ABC-Rb}G. Altarelli, R. Barbieri, and F. Caravaglios,
\NPB{405}{93}{3}.
\bibitem{ABC}G. Altarelli and R. Barbieri, \PLB{253}{90}{161}; G. Altarelli, R.
Barbieri, and S. Jadach, \NPB{369}{92}{3}; G. Altarelli, R. Barbieri, and F.
Caravaglios, \PLB{314}{93}{357}.
\bibitem{BF}M.~Boulware, D.~Finnel, \PRD{44}{91}{2054}; A.~Djouadi,
G.~Girardi, C.~Vergzegnassi, W.~Hollik and F.~Renard, \NPB{349}{91}{48}.
\bibitem{KKW}J.~D.~Wells, C.~Kolda, and G.~L.~Kane, \PLB{338}{94}{219}.
\bibitem{reviews}For a recent review see \eg, \JL, \DVN, and \AZ, Prog. Part.
Nucl. Phys. {\bf33} (1994) 303.
\bibitem{EFL} J.~Ellis, G.L.~Fogli and E.~Lisi, \PLB{333}{94}{118};
J. Erler and P. Langacker, UPR--0632T (hep-ph/9411203).
\bibitem{LargeTanB}See \eg, \JL, \DVN, X. Wang, and \AZ, \PRD{51}{95}{147}.
\bibitem{LNZI}\JL, \DVN, and A. Zichichi, \PRD{49}{94}{343}.
\bibitem{LNZII}\JL, \DVN, and A. Zichichi, \PLB{319}{93}{451}.
\bibitem{Easpects}\JL, \DVN, G. Park, X. Wang, and \AZ, \PRD{50}{94}{2164}.
\bibitem{ewcorr}\JL, \DVN, G.~T.~Park, H. Pois, and K. Yuan,
\PRD{48}{93}{3297}; \JL, \DVN, G.~T.~Park, and A. Zichichi, \PRD{49}{94}{355}.
\bibitem{eps1-epsb} \JL, \DVN, G.~T.~Park, and A. Zichichi, \PRD{49}{94}{4835}.
\bibitem{hollik}W.~Hollik, \IJMP{5}{90}{1909}; A. Denner, R. Guth, W. Hollik,
and J. K\"uhn, Z. Phys. {\bf C51} (1991) 695.
\end{thebibliography}
\end{document}